\documentclass[fleqn]{article}
\usepackage{espcrc2}
\usepackage{graphicx}
\usepackage{psfig}
\usepackage{rotating}
\usepackage{enumerate}
\usepackage{subfigure}
\newcommand{\upcite} {\cite}
\usepackage{xspace}
\usepackage{hepparticles}
\DeclareRobustCommand{\mPKst}{\HepParticle{\it K}{}{\ast 0}\xspace}%
\DeclareRobustCommand{\mAPKst}{\HepAntiParticle{\it K}{}{\ast 0}\xspace}%
\DeclareRobustCommand{\mPJpsi}{\HepParticle{\it J\mspace{-2mu}/\mspace{-2mu}\psi}{}{}\xspace}%
\DeclareRobustCommand{\mPeta}{\HepParticle{\eta}{}{}\xspace}%
\DeclareRobustCommand{\mPY}{\HepParticleResonance{Y}{2175}{}{}\xspace}

\DeclareRobustCommand{\mPKstm}{\HepParticle{\it K}{}{\ast -}\xspace}%

\begin{document}

\title{Study of $\mPJpsi$ decays into $\mPeta\mPKst\mAPKst$}

\author{\small{
M.~Ablikim$^{1}$,              J.~Z.~Bai$^{1}$,   Y.~Bai$^{1}$,
Y.~Ban$^{11}$,
X.~Cai$^{1}$,                  H.~F.~Chen$^{16}$,
H.~S.~Chen$^{1}$,              H.~X.~Chen$^{1}$, J.~C.~Chen$^{1}$,
Jin~Chen$^{1}$
% ...
\\
\vspace{0.2cm}
(BES Collaboration)\\
\vspace{0.2cm}
{\it
$^{1}$ Institute of High Energy Physics, Beijing 100049, People's Republic of China\\
% ...
$^{11}$ Peking University, Beijing 100871, People's Republic of China\\
% ...
$^{16}$ University of Science and Technology of China, Hefei 230026, 
People's Republic of China\\
% ...
}
}
}

\date{\today}

\begin{abstract}

We report the first observation of $\mPJpsi \rightarrow
\mPeta\mPKst\mAPKst$ decay in a $\mPJpsi$ sample of 58 million
events collected with the BESII detector. The branching fraction is
determined to be $(1.15 \pm 0.13 \pm 0.22)\times 10^{-3}$. The
selected signal event sample is further used to search for the $\mPY$
resonance through $\mPJpsi \rightarrow \mPeta
\mPY, \mPY\rightarrow\mPKst\mAPKst$. No evidence of a signal is seen.
An upper limit of \begin{math} \mathrm{Br}(\mPJpsi \rightarrow
\mPeta \mPY)\cdot\mathrm{Br}(\mPY\rightarrow\mPKst\mAPKst) < 2.52\times 10^{-4}
\end{math} is set at the 90\% confidence level.

\end{abstract}

\maketitle

\section{Introduction}

Following the observation of $\mPY$ by the BaBar Collaboration in
$e^+e^-\rightarrow\gamma_{ISR}\phi f_0(980)$ via initial-state
radiation~\upcite {BABAR2175}, the resonance was observed by the BES
Collaboration in $J/\psi\rightarrow\eta\phi f_0(980)$~\upcite
{BES2186} and more recently by the Belle Collaboration in
$e^+e^-\rightarrow\gamma_{ISR}\phi\pi^+\pi^- $~\upcite {BELLE2133}.
Since both the $\mPY$ and $Y(4260)$ ~\upcite{BABAR4260} are observed
in $e^+e^-$ annihilation via initial-state radiation and these two
resonances have similar decay modes, it was speculated that $\mPY$ may
be an $s$-quark version of $Y(4260)$~\upcite {BABAR2175}. There have
been a number of different interpretations proposed for the $Y(4260)$,
that include a $gc\bar{c}$ hybrid~\upcite
{Y4260h1}~\upcite{Y4260h2}~\upcite{Y4260h3}, a $4^3S_1$ $c\bar{c}$
state ~\upcite{Y4260m}, a $[cs]_S[\bar{c}\bar{s}]_S$ tetraquark state
~\upcite{Y4260t}, or a baryonium ~\upcite{Y4260b}. Likewise $\mPY$ has
been correspondingly interpreted as: a $gs\bar{s}$ hybrid
~\upcite{Y2175h}, a $2^3D_1$ $s\bar{s}$ state ~\upcite{Y2175m}, or a
$s\bar{s}s\bar{s}$ tetraquark state ~\upcite{Y2175t}. None of these
interpretations has either been established or ruled out by
experimental observations.

According to Ref.~\upcite{Y2175m}, a hybrid state may have very
different decay patterns compared to a quarkonium
state. Measuring the branching fractions of some decay modes may shed
light on understanding the nature of \mPY.  Among those promising
decay modes, $\mPY\rightarrow\mPKst\mAPKst$ is of special
importance. This decay mode is forbidden if \mPY is a hybrid state but
allowed if it is a quarkonium state.

On the other hand, there are still lots of unknown decay modes of \mPJpsi 
and investigating more of them is useful to understand the mechanism 
of \mPJpsi decays. Based on a sample of 58M \mPJpsi events collected by the BESII
detector at the Beijing Electron-Positron Collider (BEPC), a search
for the process $\mPJpsi \rightarrow \mPeta \mPY,
\mPY\rightarrow\mPKst\mAPKst$ is performed. In addition, the first
measurement of the branching fraction Br($\mPJpsi \rightarrow
\mPeta\mPKst\mAPKst$) is obtained.

\section{Detector and data samples}

The upgraded Beijing Spectrometer detector (BESII) was located at
the Beijing Electron-Positron Collider (BEPC). BESII was a large
solid-angle magnetic spectrometer which is described in detail in
Ref.~\upcite{besii}.  The momentum of charged particles is
determined by a 40-layer cylindrical main drift chamber (MDC)
which has a momentum resolution of
$\sigma_{p}$/p=$1.78\%\sqrt{1+p^2}$ ($p$ in GeV/c). Particle
identification is accomplished using specific ionization ($dE/dx$)
measurements in the drift chamber and time-of-flight (TOF)
information in a barrel-like array of 48 scintillation counters.
The $dE/dx$ resolution is $\sigma_{dE/dx}\simeq8.0\%$; the TOF
resolution for Bhabha events is $\sigma_{TOF}= 180$ ps.  Radially
outside of the time-of-flight counters is a 12-radiation-length
barrel shower counter (BSC) comprised of gas tubes interleaved
with lead sheets. The BSC measures the energy and direction of
photons with resolutions of $\sigma_{E}/E\simeq21\%/\sqrt{E}$ ($E$
in GeV), $\sigma_{\phi}=7.9$ mrad, and $\sigma_{z}=2.3$ cm. The
iron flux return of the magnet is instrumented with three double
layers of proportional counters that are used to identify muons.

A GEANT3 based Monte Carlo (MC) package (SIMBES)~\upcite{SIMBES} with
detailed consideration of real detector performance (such as dead
electronic channels) is used. The consistency between data and Monte
Carlo has been carefully checked in many high purity physics channels,
and the agreement is quite reasonable~\upcite{SIMBES}.

\section{Analysis}
The decay channel under investigation,
$\mPJpsi\rightarrow\mPeta\mPKst\mAPKst, \eta\rightarrow\gamma\gamma,
\mPKst\rightarrow K^+\pi^-, \mAPKst\rightarrow K^-\pi^+$, has two
charged kaons, two charged pions, and two photons in its final state.
A candidate event is therefore required to have four good charged
tracks reconstructed in the MDC with net charge zero and at least two
isolated photons in the BSC. A good charged track is required to (1)
be well fitted to a three dimensional helix in order to ensure a
correct error matrix in the kinematic fit; (2) originate from the
interaction region, i.e.  the point of closest approach of the track
to the beam axis is within 2 cm of the beam axis and within 20 cm from
the center of the interaction region along the beam line; (3) have a
polar angle $\theta$, within the range $|\cos\theta|<0.8$; and (4)
have a transverse momentum greater than 70 MeV/$c$. The TOF and
${\mathrm d}E/{\mathrm d}x$ information is combined to form a particle
identification confidence level for the $\pi$, $K$, and $p$
hypotheses, and the particle type with the highest confidence level is
assigned to each track. The four charged tracks selected are further
required to be consistent with an unambiguously identified $K^+
\pi^+ K^- \pi^-$ combination. An isolated neutral cluster
is considered as a good photon when (1) the energy deposited in
the BSC is greater than 60 MeV, (2) the angle between the nearest
charged track and the cluster is greater than $15^\circ$, (3) the
angle between the cluster development direction in the BSC and the
photon emission direction is less than $30^\circ$, and (4) at least
two layers have deposits in the BSC and the first hit is in the
beginning six layers. A four-constraint (4-C) kinematic fit is
performed to the hypothesis $J/\psi\rightarrow \gamma\gamma
K^+K^-\pi^+\pi^-$, and if there are more than two good photons, 
the combination with the smallest $\chi^2_{\gamma\gamma
K^+K^-\pi^+\pi^-}$ value is selected. We further require that
$\chi^2_{\gamma\gamma K^+K^-\pi^+\pi^-}<20$. Because we are not
interest in the events of which the two photons come from $\pi^0$, we
require the invariant mass of two photons to be greater than $0.3$ GeV/$c^2$.

\subsection{Branching fraction of $\mPJpsi\rightarrow\mPeta\mPKst\mAPKst$}
\begin{figure*}[htbp]
\subfigure{
\label{mkpivskpi:a}
\includegraphics[width=7.5cm,height=7.5cm]{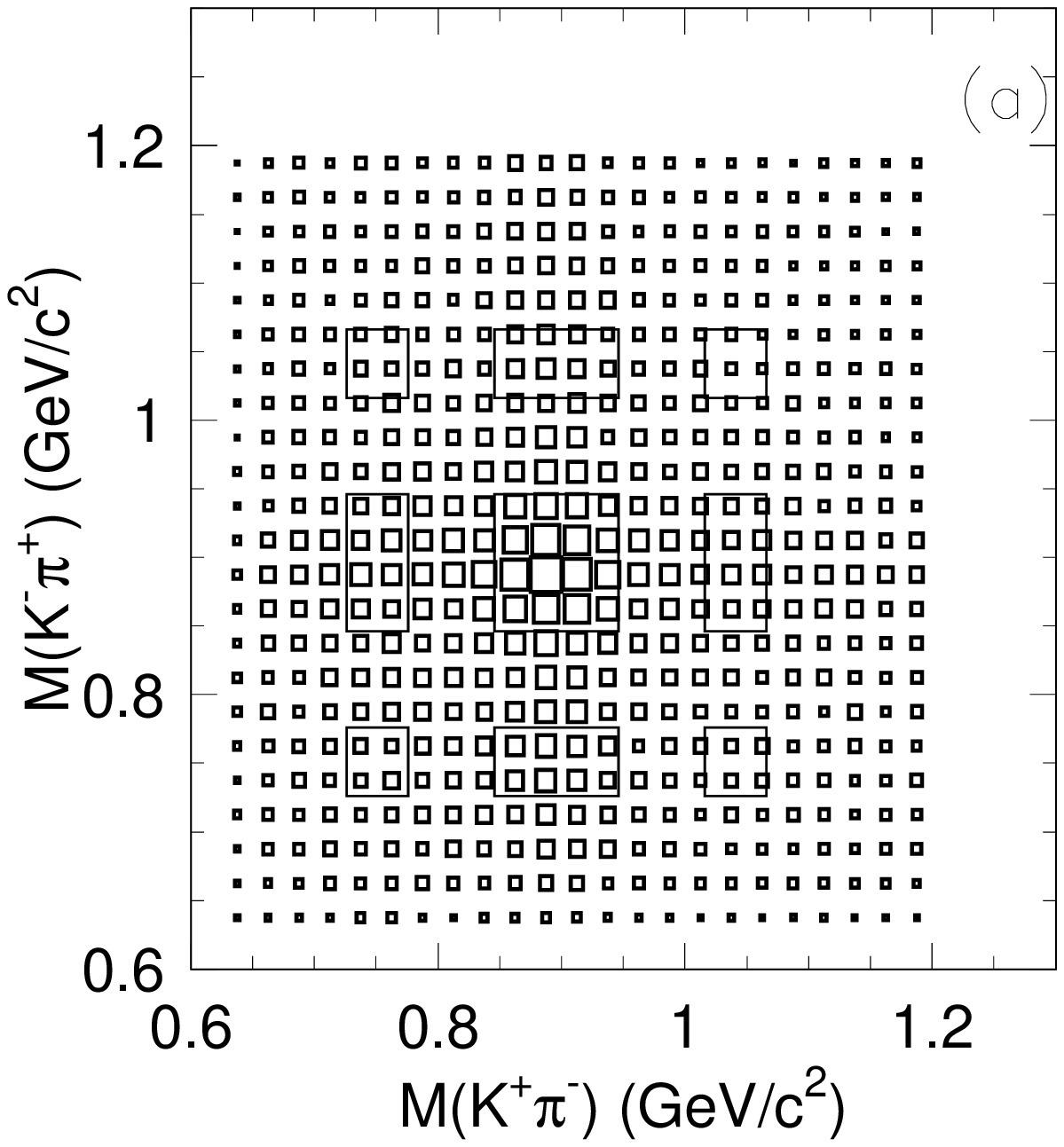}
}
\subfigure{
\label{mkpivskpi:b}
\includegraphics[width=7.5cm,height=7.5cm]{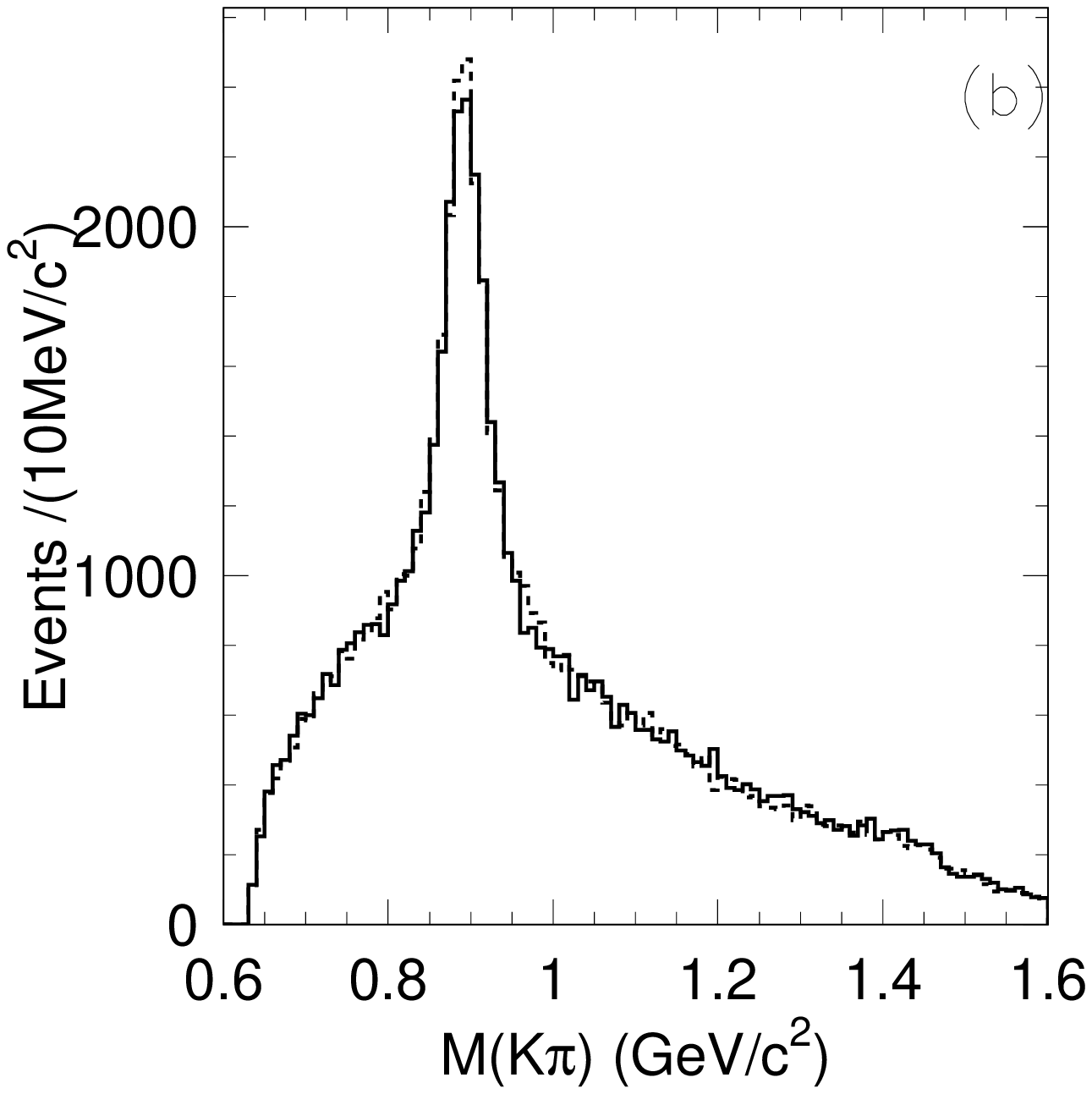}
}
\caption{(a) Scatter plot of $M_{K^+\pi^-}$ versus
  $M_{K^-\pi^+}$ invariant mass, where the middle box is the signal
  region and the other boxes are the side-band regions.  (b) The
  invariant mass spectra of $K^\pm\pi^\mp$; the solid histogram is
  $K^+\pi^-$ and the dashed is $K^-\pi^+$. }
\label{mkstvskst}
\end{figure*}

\begin{figure*}[htbp]
\subfigure{
\label{mgg2vs3}
\includegraphics[width=7.5cm,height=7.5cm]{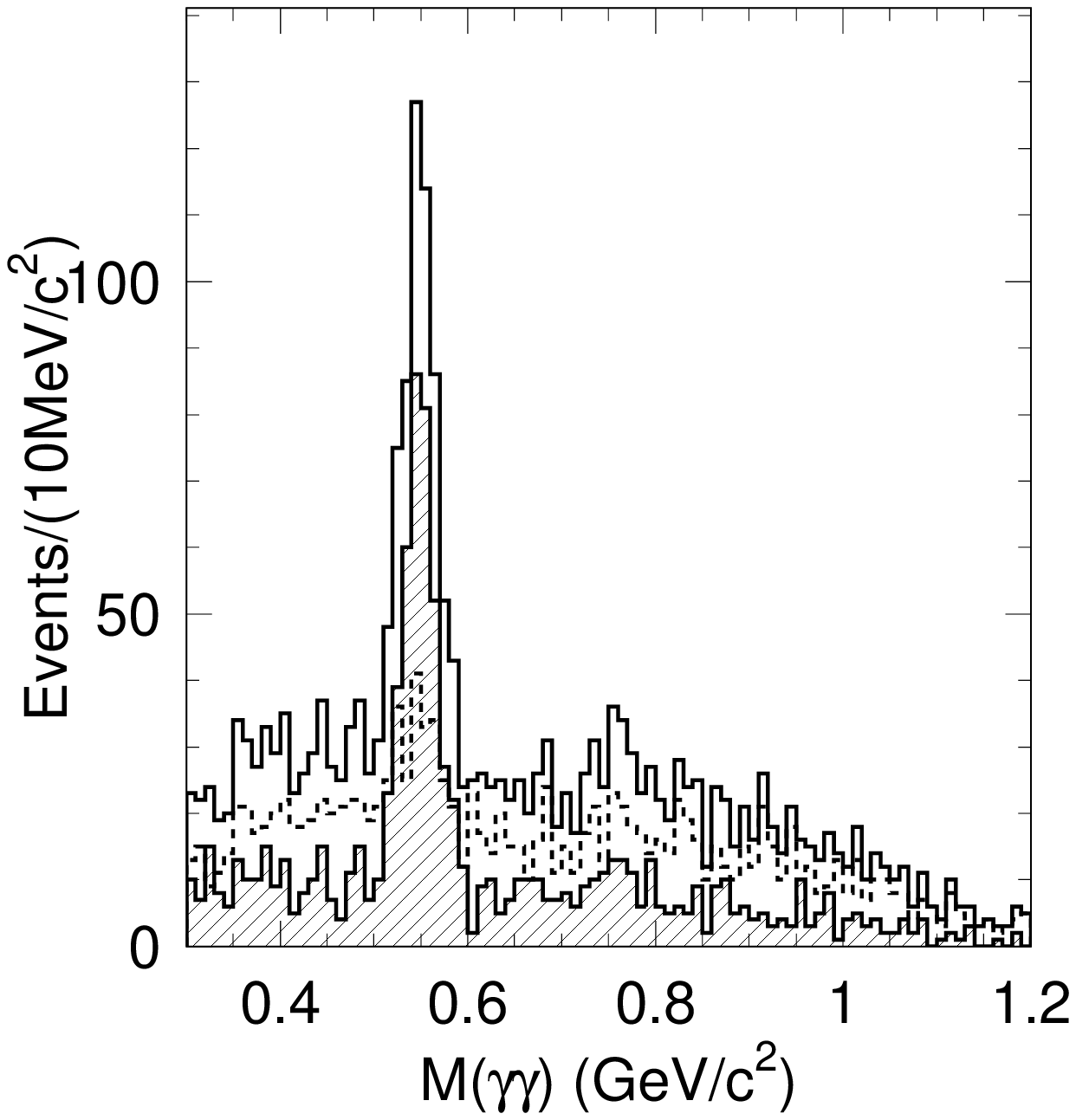}
}
\subfigure{
\label{mgg-dt-sb}
\includegraphics[width=7.5cm,height=7.5cm]{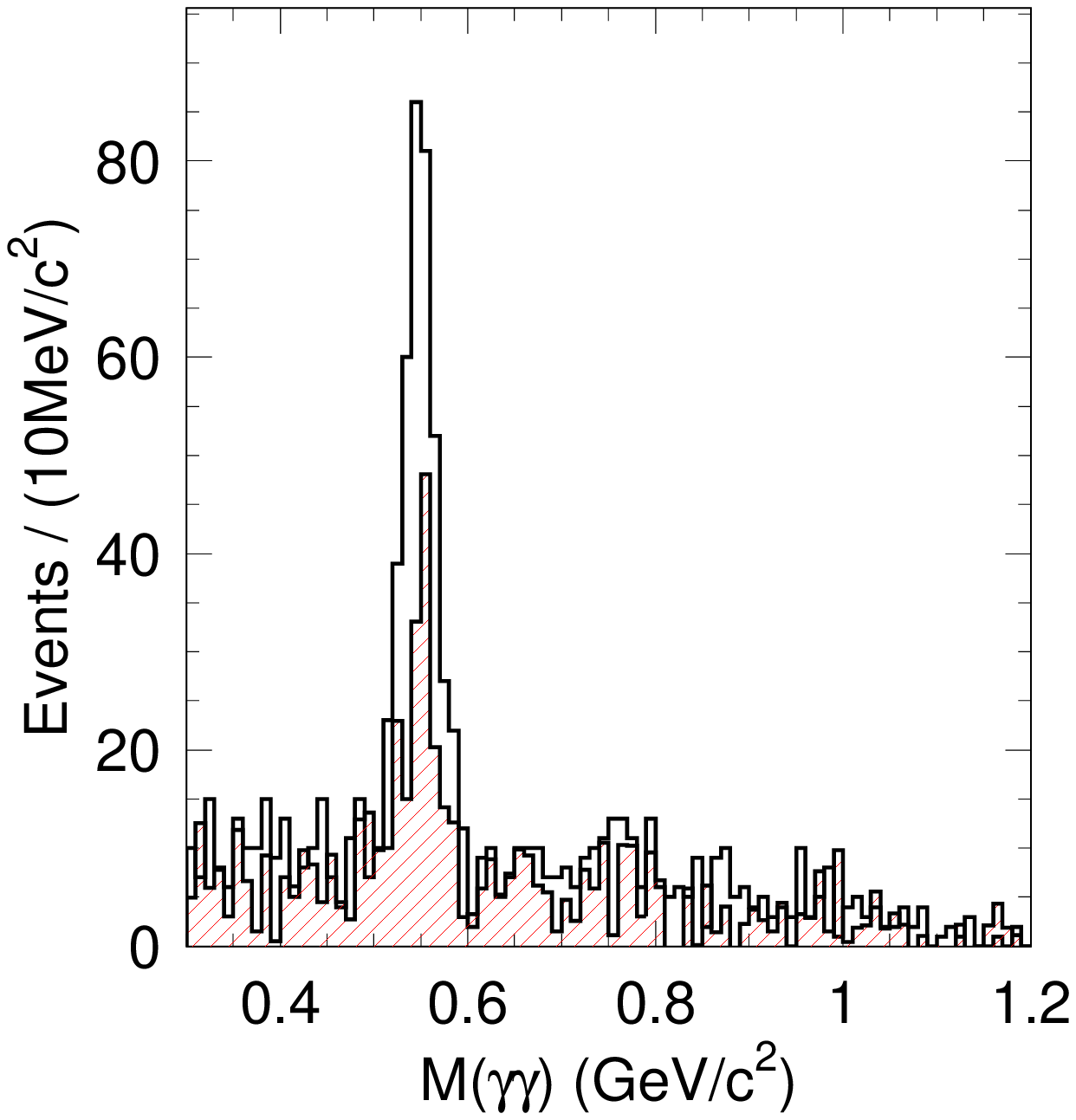}
}
\put(-300,160){\bf \large (a)}
\put(-90,160){\bf \large (b)}
\caption{(a) The $\gamma\gamma$ invariant mass spectrum for data; the
dashed histogram is from the $N_{\gamma}>2$ events, the shaded
histogram is from the $N_{\gamma}=2$ events, and the blank histogram
is from all events. (b) The $\gamma\gamma$ invariant mass spectrum for
$N_{\gamma}=2$, where the blank histogram is from signal region
events, and the shaded one is from the side-band regions events.}
\end{figure*}

After applying the above event selection criteria,
Fig.~\ref{mkpivskpi:a} shows the scatter plot of $M_{K^+\pi^-}$ versus
$M_{K^-\pi^+}$. One can see $\mPKst\mAPKst$, $\mPKst K^-\pi^+$,
$\mAPKst K^+\pi^-$, and $K^+\pi^-K^-\pi^+$ events scattered in
different regions of the plot. The signal region in this analysis is
defined by $|M_{K^\pm\pi^\mp}-m_{\mPKst}(m_{\mAPKst})|<0.05$
GeV/$c^2$, which is shown as the middle box in
Fig.~\ref{mkpivskpi:a}. Other boxes shown are side-band regions, and
events in these regions are used to estimate the background in the
signal region. The $K^\pm\pi^\mp$ invariant mass spectra are shown in
Fig.~\ref{mkpivskpi:b}, where the solid histogram is $K^+\pi^-$
and the dashed histogram is $K^-\pi^+$.  Figure \ref{mgg2vs3}
shows the $\gamma\gamma$ invariant mass spectrum for events in the signal
region, where an $\eta$ is seen. In Fig.~\ref{mgg2vs3}, the
shaded histogram is the spectrum obtained requiring two good photons,
while the dashed histogram is the spectrum for more than two
photons. When there are more than two photons, the ratio of signal
over background is much lower.  In order to remove potential
backgrounds as much as possible, we also require the number of good
photons to be two.

\begin{figure*}[htbp]
\subfigure{
\label{fitmgg:a}
\includegraphics[width=7.5cm,height=7.5cm]{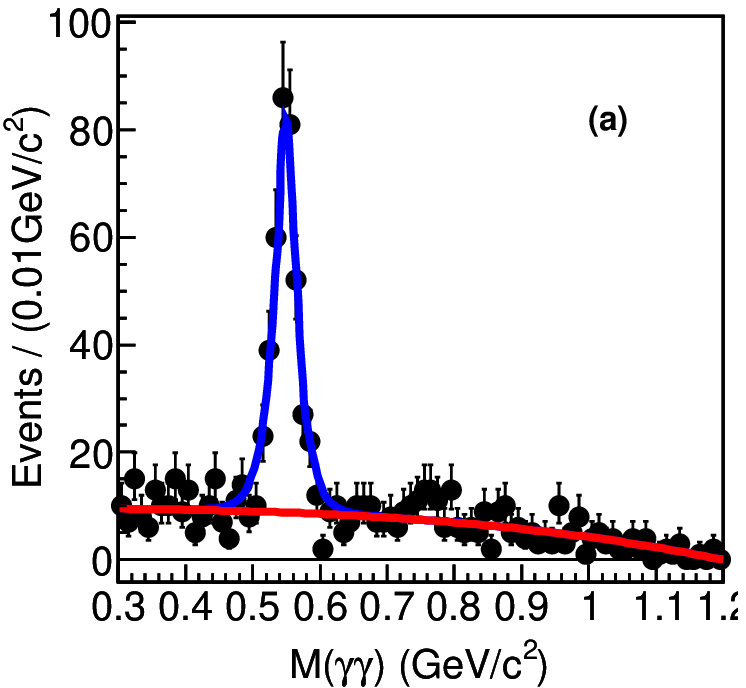}
}
\subfigure{
\label{fitmgg:b}
\includegraphics[width=7.5cm,height=7.5cm]{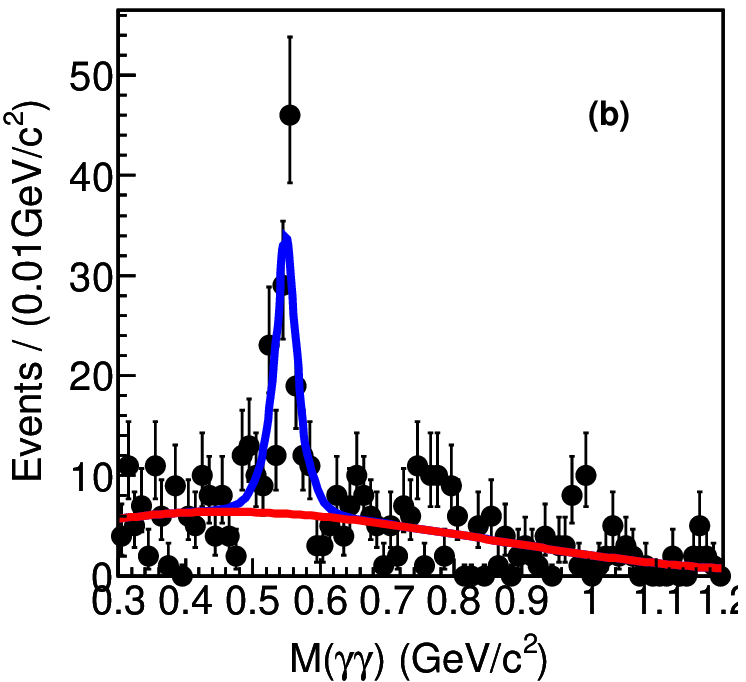}
}
\caption{Unbinned fitting results of $\gamma\gamma$ invariant mass spectra: (a) for the signal region events; (b) for the
side-band region events,
where the signal shape is obtained from the MC $\gamma\gamma$ invariant
mass distribution and the background shape is a third order Chebyshev polynomial.
}
\label{fitmgg}
\end{figure*}

Figure~\ref{mgg-dt-sb} shows the $\gamma\gamma$
invariant mass spectrum of events surviving the above selection, while
the shaded histogram is the normalized background estimated using the
side-band regions shown in Fig.~\ref{mkpivskpi:a}.  The number of
$\mPJpsi\rightarrow\mPeta\mPKst\mAPKst$ events is determined by
fitting the spectra in Fig.~\ref{mgg-dt-sb}.  The
$\mPJpsi\rightarrow\mPeta\mPKst\mAPKst$ branching fraction
is determined using
\begin{center}
\small{$Br(\mPJpsi \rightarrow \mPeta \mPKst \mAPKst) =
\frac{N_{sig}-N_{sb}}{N_{J/\psi}\cdot \epsilon\cdot Br(\mPKst\rightarrow K^+\pi^-)
\cdot Br(\mAPKst\rightarrow
K^-\pi^+)\cdot Br(\mPeta\rightarrow\gamma\gamma)}$,}
\end{center}
where $N_{sig}=347$ is the number of events in the signal region,
obtained by fitting the spectrum in Fig.~\ref{mgg-dt-sb} (the blank
histogram); $N_{sb}=138$ is the number of background events estimated from
side-band regions, obtained by fitting the spectrum in
Fig.~\ref{mgg-dt-sb} (the shaded histogram); $N_{\mPJpsi}$ is the
total number of $\mPJpsi$ events ~\upcite{jpsinum}; $\epsilon=1.79\%$ is the
detection efficiency obtained from MC simulation of
$\mPJpsi\rightarrow\mPeta \mPKst \mAPKst$; and $Br(\mPKst\rightarrow
K^+\pi^-)$, $Br(\mAPKst\rightarrow K^-\pi^+)$ and
$Br(\mPeta\rightarrow\gamma\gamma)$ are the corresponding branching
fractions.  Figures \ref{fitmgg:a} and \ref{fitmgg:b} show
respectively the fitting results of the signal and side-band events,
where the shape of the $\gamma\gamma$ invariant mass spectrum obtained
from the MC sample $\mPJpsi\rightarrow\mPeta \mPKst \mAPKst$ is used
as the signal shape and a third order Chebyshev polynomial is used as
the background shape.  The $\mPJpsi\rightarrow\mPeta \mPKst \mAPKst$ branching
fraction is determined to be
\begin{center}
\small{$Br(\mPJpsi \rightarrow \mPeta \mPKst \mAPKst) =
(1.15\pm 0.13)\times 10^{-3},$}
\end{center}
where the error is statistical only. It is the first measurement for this decay mode of \mPJpsi 
and it is shown that this mode is a typical three bodies decay. The branching fraction is compatible 
with the result of $Br(\mPJpsi\rightarrow\mPeta K^+K^-\pi^+\pi^-)=(1.84\pm 0.28)\times 10^{-3}$ given by
BaBar Collaboration ~\upcite{BaBar_br}. 
It is worth mention of that this branching fraction is several times smaller 
than the radiative decay mode $\mPJpsi\rightarrow\gamma \mPKst \mAPKst$ which is very different from
the situation of $p\bar p$ that the branching fraction of $\mPJpsi\rightarrow\mPeta p \bar p$ is much bigger than 
$\mPJpsi\rightarrow\gamma p \bar p$.

\subsection{$\mPJpsi\rightarrow\mPeta \mPY\rightarrow\mPeta\mPKst\mAPKst $} 
\begin{figure*}[htbp]
\centering
\includegraphics[width=7.5cm,height=7.5cm]{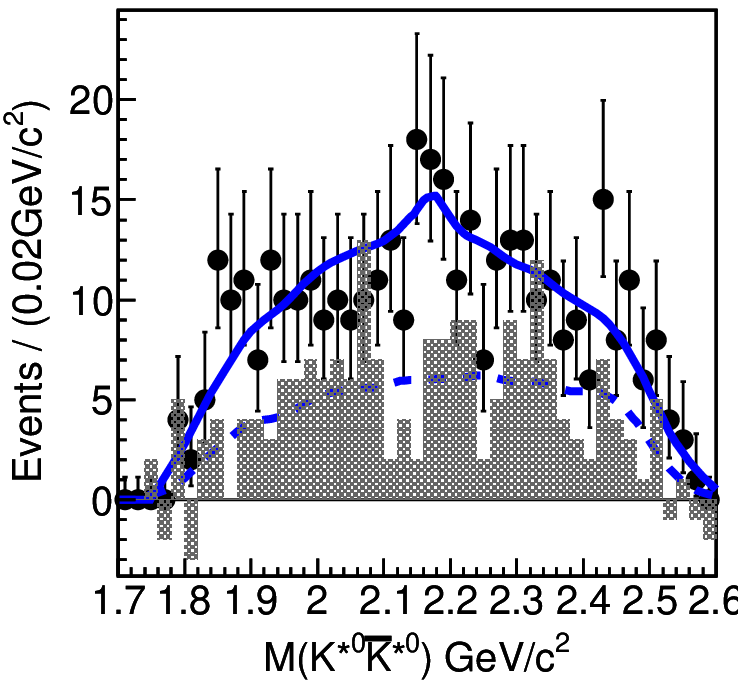}
\caption{The $\mPKst\mAPKst$ invariant mass spectrum, where points with
  error bars are candidate events, the
  dashed histogram is from MC phase space for
  $\mPJpsi\rightarrow\mPeta\mPKst\mAPKst$, the shaded histogram is from
  side-band events, and the solid curve is the fitting result, where
  the Y(2175) shape used is from MC simulation.}
\label{mkstkst_fit}
\end{figure*}

Next, we search for a possible resonance recoiling against
$\mPeta$.  So in addition to the above requirements, we require that
the $\gamma\gamma$ invariant mass satisfies
$|M_{\gamma\gamma}-m_{\mPeta}|<0.04$ GeV/$c^2$ and define the
side-band region to be 0.1 GeV/$c^2$ $<|M_{\gamma\gamma}-m_{\mPeta}|<0.14$
GeV/$c^2$. The $\mPKst\mAPKst$ invariant mass spectrum recoiling
against $\mPeta$ for
$\mPJpsi\rightarrow\mPeta\mPKst\mAPKst$ is
shown in Fig.~\ref{mkstkst_fit}, where the dashed histogram is the
contribution from phase space for
$\mPJpsi\rightarrow\mPeta\mPKst\mAPKst$ and the shaded histogram is
the contribution from the normalized side-band events in the $\mPeta$,
$\mPKst$ and $\mAPKst$ side-band
regions.  There is
no obvious enhancement in the region around 2.175 GeV/$c^2$.

The backgrounds in the selected event sample are studied with MC
simulations.  For the decay $\mPJpsi\rightarrow\mPeta\mPKst\mAPKst$,
the possible main background channels are:
$\mPJpsi\rightarrow\eta\mPKst\mAPKst\rightarrow
(3\pi^0)\mPKst\mAPKst$; $\mPJpsi\rightarrow a_0^+ K^-
\mPKst\rightarrow (\eta \pi^+)K^- \mPKst + c.c.$;
$\mPJpsi\rightarrow\rho^+ \mPKstm \mPKst\rightarrow (\pi^+ \pi^0)(K^-
\pi^0) \mPKst + c.c.$; $\mPJpsi\rightarrow\gamma \pi^0\mPKst\mAPKst$;
$\mPJpsi\rightarrow\phi \eta'\rightarrow K^+K^-\eta\pi^+\pi^-$; for
each channel a sizable MC sample is simulated. There is no peak around
2.175 GeV/$c^2$ in the $\mPKst\mAPKst$ invariant mass distribution in any
background channel.

We fit the mass distribution to determine a possible signal, where
three parts are included in the total probability distribution
function (p.d.f): (1) for the signal p.d.f, we use the shape of the
$\mPKst\mAPKst$ invariant mass spectrum obtained from MC simulation of
$\mPJpsi\rightarrow\mPeta\mPY\rightarrow\mPeta\mPKst\mAPKst$ produced
with the mass and width of $\mPY$ fixed to BaBar's results; (2) for
the normalized phase space contribution p.d.f., we use the shape of
the $\mPKst\mAPKst$ invariant mass distribution obtained in the
$\mPJpsi\rightarrow\mPeta\mPKst\mAPKst$ MC simulation, normalized with
the branching ratio obtained in the previous section; (3) for the
other possible backgrounds, we use a third order Chebyshev polynomial.

The product branching ratio is determined using
\begin{center}
\small{$Br(\mPJpsi\rightarrow\mPeta\mPY)\cdot Br(\mPY\rightarrow\mPKst\mAPKst)
=\frac{N^{obs}}{N_{\mPJpsi}\cdot \epsilon\cdot Br(\mPKst\rightarrow K^+\pi^-)
\cdot Br(\mAPKst\rightarrow
K^-\pi^+)\cdot Br(\mPeta\rightarrow\gamma\gamma)}
=(0.7\pm 0.8)\times 10^{-4},$} 
\end{center}
where $N^{obs} = 11\pm 12$ is the number of signal
events, $N_{\mPJpsi}$ is the total number of $\mPJpsi$ events
~\upcite{jpsinum}, $\epsilon=1.57\%$ is the detection efficiency obtained
from MC simulation of
$\mPJpsi\rightarrow\mPeta\mPY\rightarrow\eta\mPKst\mAPKst$, where the
first step decay used an angular distribution $1+\cos^2\theta$,
$\theta$ is the polar angle of the $\mPeta$ momentum in the center of
mass frame, $Br(\mPKst\rightarrow K^+\pi^-)$ and
$Br(\mAPKst\rightarrow K^-\pi^+)$ and
$Br(\mPeta\rightarrow\gamma\gamma)$ are the corresponding branching
fractions.  The error is only the statistical error. The signal
significance is only $0.88\sigma$.

The upper limit of
$Br(\mPJpsi\rightarrow\mPeta \mPY)\cdot Br(\mPY\rightarrow
\mPKst\mAPKst)$ at the 90\% confidence level is obtained using a Bayesian
approach ~\upcite{PDG08}.  
We obtain the upper limit: 
\begin{center}
\small{$Br(\mPJpsi\rightarrow\mPeta\mPY)\cdot Br(\mPY\rightarrow\mPKst\mAPKst) 
<\frac{N^{obs}_{up}}{N_{\mPJpsi}\cdot \epsilon\cdot Br(\mPKst\rightarrow K^+\pi^-)
\cdot Br(\mAPKst\rightarrow
K^-\pi^+)\cdot Br(\mPeta\rightarrow\gamma\gamma)\cdot
(1-\sigma^{sys})}
= 2.52\times 10^{-4}, $}
\end{center}
where $N^{obs}_{up}=31$ is upper limit at the 90\% confidence level,
$\sigma^{sys}$ is the systematic error discussed below, and the other
symbols are defined as above.

\section{Systematic Errors}
In this analysis, the systematic errors on the branching fraction and
upper limit mainly come from the following sources:

\subsection{MDC Tracking efficiency and kinematic fitting}
The systematic errors from MDC tracking and kinematic fitting are
estimated by using simulations with different MDC wire resolutions
~\upcite{SIMBES}. In this analysis, the systematic errors from this
source are 12.8\% for $\mPJpsi\rightarrow\mPeta\mPKst\mAPKst$ and
12.0\% for
$\mPJpsi\rightarrow\mPeta\mPY\rightarrow\mPeta\mPKst\mAPKst$.

\subsection{Photon detection efficiency}
The photon detection efficiency is studied in reference~\upcite{SIMBES}.
The results indicate that the systematic error is less than 1\% for
each photon. Two good photons are required in this analysis, so 2\% is
taken as the systematic error for the photon detection efficiency.

\subsection{Particle identification (PID)}
In references~\upcite{SIMBES} and \upcite{KPID}, the efficiencies of pion
and kaon identification are analyzed. The systematic error from PID
is about 1\% for each charged track. In this analysis, four charged
tracks are required, so 4\% is taken as the systematic error from
PID.

\subsection{Uncertainty of intermediate decay}
The branching fraction uncertainties for $\mPeta \rightarrow
\gamma\gamma$ and $\mPKst(\mAPKst)\rightarrow K^+\pi^-(K^-\pi^+)$ from
PDG08 ~\upcite{PDG08} are taken as systematic errors.

\subsection{Number of $\mPJpsi$ events}
The number of $\mPJpsi$ events is $(57.70 \pm 2.62) \times 10^6$,
determined from the number of inclusive 4-prong
hadrons~\upcite{jpsinum}. The uncertainty 4.72\% is taken as a
systematic error.

\subsection{Fitting}
\subsubsection{$\mPJpsi \rightarrow
\mPeta\mPKst\mAPKst$ branching fraction}
When fitting the $\gamma\gamma$ invariant mass spectrum, as described
in section III.A, the $\mPeta$ signal shape obtained from MC is fixed,
and different order polynomials are used for the background shape. The
difference is taken as the systematic error for the background
uncertainty. We also use different
regions in fitting the invariant mass spectrum. The total systematic
error from fitting is 6.7\%.

\subsubsection{$\mathrm{Br}(\mPJpsi \rightarrow
\mPeta \mPY)\cdot\mathrm{Br}(\mPY\rightarrow\mPKst\mAPKst)$ upper limit}

When fitting the invariant mass spectrum of $\mPKst\mAPKst$, as
described in section III.B, there are three sources of systematic
error: for the first p.d.f, we used the different resonance parameters
measured by BaBar and BES, and take the difference as the systematic
error from the uncertainty of signal parameters; for the second, the
systematic error comes from the error of the branching fraction of
$\mPJpsi\rightarrow\mPeta\mPKst\mAPKst$ measured in section III.A; for
the third, we used the difference between fitting with a third order
Chebyshev polynomial and fitting with the invariant mass shape from
$\mPKst\mAPKst$ side-band events as the systematic error for the
background uncertainty. Combining these contributions,
16.3\% is obtained as the systematic error from fitting.

\subsection{Different selection of side-band regions}
We used different side-band regions to estimate the backgrounds both
in section III.A and III.B, and take the difference as a source of
systematic error. The result is 10.0\% for the measurement of
branching fraction and 4.2\% for the upper limit.

\subsection{Number of photons}
To estimate the systematic error from the requirement of two good
photons, we compare the efficiency difference for this requirement
between data and MC sample, and obtain 4.4\%, which is taken as the
systematic error from the two photon requirement.

\subsection{$K^*$ simulation}
The $K^*$ is simulated with a P-wave relativistic Breit-Wigner
function
$BW=\frac{{\Gamma(s)}^2{m_0}^2}{(s-{m_0}^2)^2+{\Gamma(s)}^2{m_0}^2}$,
with the width
$\Gamma(s)=\Gamma_0\frac{m_0}{m}\frac{1+r^2{p_0}^2}{1+r^2p^2}[\frac{p}{p_0}]^3$,
where r is the interaction radius and the value
$(3.4\pm0.6\pm0.3)(GeV/c)^{-1}$ measured by a $K^-\pi^+$ scattering
experiment ~\upcite {kstpwave} is used. Varying the value of r by
$1\sigma$, the difference of the detection efficiencies for
$\mPJpsi\rightarrow\eta\mPKst\mAPKst$,
$\mPJpsi\rightarrow\mPeta\mPY\rightarrow\mPeta\mPKst\mAPKst$ is taken
as the systematic error from the uncertainty of the r value.

The systematic errors from the different sources and the total
systematic errors are shown in Table I.
\begin{table*}[htpb]
\begin{center}
\caption{Systematic errors (\%)}
\begin{tabular}{l|c|c}
\hline \hline Error sources & Br($\mPJpsi\rightarrow\mPeta\mPKst\mAPKst$) & Upper limit \\
\hline MDC tracking efficiency and 4-C fitting & 12.8 & 12.0 \\
\hline Photon detection efficiency & 2 & 2 \\
\hline PID & 4 & 4 \\
\hline Intermediate decay & $\sim 1$ & $\sim 1$ \\
\hline Number of $J/\psi$ events & 4.7 & 4.7 \\
\hline Fitting & 6.7 & 16.3 \\
\hline Side-band region & 10.0 & 4.2 \\
\hline Photon number & 4.4 & 4.4 \\
\hline $K^*$ simulation & 3.5 & 2.6 \\
\hline Total systematic error & 19.5 & 22.3 \\
\hline
\end{tabular}
\end{center}
\end{table*}

\section{Summary}
With 58M BESII $\mPJpsi$ events, the branching fraction of
$\mPJpsi\rightarrow\mPeta\mPKst\mAPKst$ is measured for the first time:
\begin{center}
\small{$Br(\mPJpsi\rightarrow\mPeta\mPKst\mAPKst)=(1.15\pm0.13\pm0.22)\times 10^{-3}.$}
\end{center}

No obvious enhancement near 2.175 GeV/$c^2$ in the invariant mass
spectrum of $\mPKst\mAPKst$ is observed. The upper limit on
$Br(\mPJpsi\rightarrow\mPeta\mPY)\cdot Br(\mPY\rightarrow
\mPKst\mAPKst)$ at the 90\% C.L. is $2.52\times 10^{-4}$. Due to the
limited statistics, we can not distinguish whether the $\mPY$ is a
hybrid or quarkonium state.


The BES collaboration thanks the staff of BEPC and computing
center for their hard efforts. This work is supported in part by
the National Natural Science Foundation of China under contracts
Nos. 10491300, 10225524, 10225525, 10425523, 10625524, 10521003, 10821063,
10825524, the Chinese Academy of Sciences under contract No. KJ 95T-03, the
100 Talents Program of CAS under Contract Nos. U-11, U-24, U-25,
and the Knowledge Innovation Project of CAS under Contract Nos.
U-602, U-34 (IHEP), the National Natural Science Foundation of
China under Contract No. 10225522 (Tsinghua University), and the
Department of Energy under Contract No. DE-FG02-04ER41291 (U.
Hawaii).

\end{document}